\documentstyle[12pt]{article}

\makeatletter
\@addtoreset{equation}{section}
\makeatother

\def\half{{\textstyle{1\over2}}}
\def\dalemb#1#2{{\vbox{
        \hrule height .#2pt
        \hbox{\vrule width.#2pt height#1pt \kern#1pt \vrule width.#2pt}
        \hrule height.#2pt}}}

\let\la=\label  
  
 \def\bd{\begin{document}} \def\ed{\end{document}}
\def\ds{\documentstyle} \let\fr=\frac \let\bl=\bigl \let\br=\bigr
\let\Br=\Bigr \let\Bl=\Bigl
\let\bm=\bibitem
\let\na=\nabla
\let\pa=\partial \let\ov=\overline
\newcommand{\be}{\begin{equation}}
\newcommand{\ee}{\end{equation}}
\def\ba{\begin{array}}
\def\ea{\end{array}}
\newcommand{\ho}[1]{$\, ^{#1}$}
\newcommand{\hoch}[1]{$\, ^{#1}$}
\newcommand{\bea}{\begin{eqnarray}}
\newcommand{\eea}{\end{eqnarray}}
\newcommand{\ra}{\rightarrow}
\newcommand{\lra}{\longrightarrow}
\newcommand{\Lra}{\Leftrightarrow}
\newcommand{\ap}{\alpha^\prime}
\newcommand{\bp}{\tilde \beta^\prime}
\newcommand{\tr}{{\rm tr} }
\newcommand{\Tr}{{\rm Tr} }
\newcommand{\NP}{Nucl. Phys. }
\newcommand{\tamphys}{\it
Center for Theoretical Physics, Department of Physics\\
Texas A\&M University, College Station, Texas 77843--4242}

\begin{document}

\rightline{CTP-TAMU-03/99}
\rightline{RU99-1-B}
\rightline{hep-th/9901149}

\vspace{24pt}

\begin{center}
{\large {\bf Anti-de Sitter Black Holes in Gauged $N=8$ Supergravity}}

\vspace{24pt}

M.~J.~Duff${}^a$\footnote{
Research supported in part by NSF Grant PHY-9722090.}
and James T.~Liu${}^b$\footnote{
Research supported in part by
the U.~S.~Department of Energy under grant no.~DOE-91ER40651-TASKB.}

\vspace{10pt}

${}^a$ {\tamphys}

\bigskip

${}^b$ {\it Department of Physics, The Rockefeller University\\
1230 York Avenue, New York, NY 10021-6399}

\bigskip

\vspace{24pt}

\underline{ABSTRACT}

\end{center}

We present new anti-de Sitter black hole solutions of gauged
$N=8$, $SO(8)$ supergravity, which is the massless sector of the
$AdS_{4}\times S^{7}$ vacuum of $M$-theory.
By focusing on the $U(1)^{4}$ Cartan subgroup, we find non-extremal
1, 2, 3 and 4 charge solutions. In the extremal limit, they may
preserve up to 1/2, 1/4, 1/8 and 1/8 of
the supersymmetry, respectively. In the limit of vanishing $SO(8)$
coupling constant, the solutions reduce to the familiar black holes
of the $M_{4}\times T^{7}$ vacuum, but have very different
interpretation since there are no winding states on $S^{7}$ and no
$U$-duality. In contrast to the $T^{7}$ compactification, moreover, we
find no static multi-center solutions. Also in contrast, the
$S^{7}$ fields appear ``already dualized'' so that the 4 charges may be all
electric or all magnetic rather than 2 electric and 2 magnetic.
Curiously, however, the magnetic solutions preserve no supersymmetries.
We conjecture that a subset of the extreme electric black holes preserving
1/2 the supersymmetry may be identified with the $S^{7}$ Kaluza-Klein
spectrum, with the non-abelian $SO(8)$ quantum numbers provided by the
fermionic zero modes.

\vfill
\leftline{}

\newpage


\section{Introduction}

The correspondence between anti-de Sitter space and conformal field
theories on its boundary \cite{Maldacena,Gubser,Witten,Witten2} has
revived an interest in gauged
extended supergravities which arise as the massless sector of the
Kaluza-Klein compactifications of $D=11$
supergravity, such as $AdS_{4}\times S^{7}$ and $AdS_{7}\times S^{4}$
or Type $IIB$ supergravity, such as $AdS_{5}\times S^{5}$ \cite{Duffads}.
Gauged $N=8$ $D=4$
supergravity \cite{deWit1,deWit2}, which is the massless sector of the $S^{7}$
compactification, has also featured in
a recent cosmological context with attempts to
reconcile an open universe with inflation
\cite{Hawkingturok,Bousso,Bremer,Hawking}.
Although this Kaluza-Klein compactification was the subject of much
investigation in the past \cite{Duffnilssonpope}, relatively little effort
has been devoted to the issue of black hole solutions of the gauged
$N=8$ theory%
\footnote{BPS black holes arising in the $SU(2) \times
SU(2)$ version of gauged $(N=4,D=4)$ supergravity, which is the
massless sector of the $S^{3} \times S^{3}$ compactification of $(N=1,D=10)$
supergravity, have recently been discussed in \cite{Klemm}. Our solutions
will be significantly different from these. In particular ours are
asymptotically $AdS$ while those of \cite{Klemm} are asymptotically neither
$AdS$ nor Minkowski.  Additionally, both BPS and non-BPS black hole
solutions in gauged $(N=2,D=5)$ supergravity were examined in
\cite{Behrndt1} and \cite{Behrndt2} respectively.}.
This is the subject of the present paper.

Although both are maximally symmetric, the $S^{7}$ compactification differs
from the $T^{7}$ in several important respects.  First of all, the global
$SO(8)$ is promoted to a gauge symmetry.  Secondly, the underlying
supersymmetry algebra is no longer Poincare but rather $AdS_4$ and the
Lagrangian has a non-vanishing cosmological constant $\Lambda$ proportional
to the square of the gauge coupling constant $g$:
\be
G\Lambda \sim -g^{2},
\la{G}
\ee
where $G$ is Newton's constant. Consequently, we shall be seeking
black hole solutions that are asymptotically $AdS$
rather than Minkowski. We also face the difference that the gauge
group is non-abelian.
By focusing on the $U(1)^{4}$ Cartan subgroup, we find
non-extremal 1, 2, 3 and 4 charge solutions. In the extremal limit
they may preserve up to 1/2, 1/4, 1/8 and 1/8 of
the supersymmetry, respectively. In the limit of vanishing $SO(8)$
coupling constant, the solutions reduce to the familiar black holes
of the $M_{4}\times T^{7}$ vacuum, but have very different
interpretation since there are no winding states on $S^{7}$ and no
$U$-duality. In contrast to the $T^{7}$ compactification, moreover, we
find no static multi-center solutions. Also in contrast, the
$S^{7}$ fields appear ``already dualized'' so that the 4 charges may be all
electric or all magnetic rather than 2 electric and 2 magnetic.
Curiously, however, the magnetic solutions preserve no supersymmetries.

Previous papers \cite{Rahmfeld1,Kaluza,Khuri} have explored the possibility
that the BPS spectrum of toroidally compactified string and M theory, and in
particular the Kaluza-Klein
modes, could be identified with extreme black hole solutions of the
low-energy supergravity theory. It was found that this identification
was consistent not only with the mass and
charge spectrum \cite{Rahmfeld1} but also with the spins and supermultiplet
structure
\cite{Duffliurahmfeld1,Rahmfeld2,Rahmfeld3} and with the dipole moments and
gyromagnetic ratios
\cite{Duffliurahmfeld2}.  In a similar spirit, we here conjecture that
a subset of the
$AdS$ electric black hole solutions preserving half the supersymmetry may be
identified
with the $S^{7}$ Kaluza-Klein spectrum
\cite{Biran,Sezgin,Duffnilssonpope} with the non-abelian $SO(8)$ quantum
numbers provided by the fermionic zero modes.

\section{$N=8$ gauged supergravity}

We follow the conventions of \cite{deWit1,deWit2}, and denote
the fields of the massless $N=8$ supergravity multiplet by
$(e_\mu^\alpha,\psi_\mu^i,A_\mu^{[IJ]},\chi^{[ijk]},{\cal
V}_{[ij]}{}^{[IJ]})$, where $i,j$ are $SU(8)$ indices and $I,J$ are $SO(8)$
indices.  The 70 real scalar degrees of freedom are represented by the
56-bein
\begin{equation}
{\cal V}=\pmatrix{u_{ij}{}^{IJ}&v_{ijKL}\cr
v^{klIJ}&u^{kl}{}_{KL}},
\end{equation}
transforming under local $SU(8)$ and rigid $E_7$.  In the gauged
supergravity theory, the 28 vectors $A_\mu^{IJ}$ transform in the adjoint
of $SO(8)$, with resulting non-abelian field strengths
$F_{\mu\nu}^{IJ}=2(\partial_{[\mu}A_{\nu]}^{IJ}-gA_{[\mu}^{IK}A_{\nu]}^{KJ})$.
We also define the fully $SO(8)\times SU(8)$ covariant derivative as, for
example, $D_\mu\varphi_i{}^I=\nabla_\mu\varphi_i{}^I
-\half{\cal B}^j_{\mu\,i}\varphi_j{}^I-g A_\mu^{IJ}\varphi_i{}^J$.  Here
${\cal B}^i_{\mu\,j}$ is a composite $SU(8)$ connection, defined along with
the scalar kinetic terms ${\cal A}_\mu^{ijkl}$ according to the condition
\begin{equation}
D_\mu{\cal V}{\cal V}^{-1}=-{1\over2\sqrt2}\pmatrix{
0&{\cal A}_\mu^{ijkl}\cr
{\cal A}_{\mu\,ijkl}&0}.
\label{eq:vdv}
\end{equation}
Note that here $D_\mu$ is the fully covariant derivative, so that
${\cal B}^i_{\mu\,j}$ is defined indirectly by the vanishing of the diagonal
blocks in (\ref{eq:vdv}).

While the complete gauged $N=8$ Lagrangian is rather involved
\cite{deWit2}, the bosonic part is fairly standard, and may be
written as
\begin{equation}
{\cal L}={1\over2\kappa^2}\sqrt{-g}\left[R
-{1\over2\cdot4!} {\cal A}_\mu^{ijkl}{\cal A}^\mu_{ijkl}
-{1\over4}(F_{\mu\nu\,IJ}^+(2S^{IJ,KL}-\delta^{IJ}_{KL})
F^{+\,\mu\nu}{}_{KL}+{\rm h.c.})-V\right],
\label{eq:lag}
\end{equation}
where $S^{IJ,KL}$ is defined in terms of the scalars through the condition
$(u^{ij}{}_{IJ}+v^{ijIJ})S^{IJ,KL}=u^{ij}{}_{KL}$, and $F_{\mu\nu}^+$ is the
self-dual part of $F_{\mu\nu}$.  Finally, the potential arises from the
$SO(8)$ gauging, and is given by
\begin{equation}
V=-2g^2[{\textstyle {3\over4}}|A_1{}^{ij}|^2-{\textstyle{1\over24}}
|A_{2\,i}{}^{jkl}|^2],
\end{equation}
where $A_1^{ij}={4\over21}T_k{}^{ikj}$ and $A_{2\,i}{}^{jkl}=-{4\over3}
T_i{}^{[jkl]}$ and $T_i{}^{jkl}$ is the $T$-tensor of \cite{deWit2}:
\begin{equation}
T_i{}^{jkl}=(u^{kl}{}_{IJ}+v^{klIJ})
(u_{im}{}^{JK}u^{jm}{}_{KI}-v_{imJK}v^{jmKI}).
\end{equation}

In a purely bosonic background, the supersymmetry transformations of the
fermions are given by
\begin{eqnarray}
\half\delta\psi_\mu^i&=&D_\mu\epsilon^i+
{\textstyle{1\over\sqrt{2}}} [
{\textstyle{1\over4}}\overline{F}_{\nu\lambda}^{-\,ij}
\gamma^{\nu\lambda}-gA_1^{ij}] \gamma_\mu
\epsilon_j,\nonumber\\
\delta\chi^{ijk}&=&-\half {\cal A}_\mu^{ijkl}\gamma^\mu\epsilon_l
+[{\textstyle{3\over2}}\gamma^{\mu\nu}\overline{F}_{\mu\nu}^{-\,[ij}
\delta^{k]}_l-2gA_{2\,l}{}^{ijk}]\epsilon^l,
\label{eq:fsusy}
\end{eqnarray}
where $D_\mu\epsilon^i=\nabla_\mu\epsilon^i+\half{\cal
B}^i_{\mu\,j}\epsilon^j$.  Note that the gauge fields enter the
supersymmetry transformations with scalar factors, since (to lowest order)
$\overline{F}_{\mu\nu}$ is defined through $F_{\mu\nu}{}^{IJ}=
(u_{ij}{}^{IJ}+v_{ijIJ})\overline{F}_{\mu\nu}{}^{ij}$.

In the case of ungauged $N=8$ supergravity, black holes are completely
characterized by 28 electric and 28 magnetic charges under the $U(1)$ gauge
fields.  In the present case, however, the gauge group is non-abelian, and
hence the situation is less clear.  In order to proceed, we work in
an abelian truncation of the gauged $N=8$ theory by focusing only on the
$U(1)^4$ Cartan subgroup of $SO(8)$.  In particular, we choose the Cartan
generators to correspond to adjacent index pairs:
\begin{equation}
\{A_\mu^{12},\quad A_\mu^{34},\quad
A_\mu^{56},\quad A_\mu^{78}\},
\end{equation}
and set the remaining gauge fields to zero.  Note that while in principle it
is important to check that this provides a consistent truncation, in
practice as long as the supersymmetry variations (at least partially) vanish
the state is essentially ensured to be BPS.

For the scalars, we work in symmetric gauge
\cite{Cremmer1,Cremmer2} where the 56-bein may be written as
\begin{equation}
{\cal V}=\exp\left\{-{1\over2\sqrt{2}}\pmatrix{0&\phi_{ijkl}\cr
\phi^{mnpq}&0}\right\},
\end{equation}
with $\phi^{ijkl}$ self-dual. Let us denote
$SO(8)$ index pairs $\{12,34,56,78\}$ by $(\alpha)$ where
$\alpha=1,\ldots,4$. Specializing
to real scalars, we are lead to the following ansatz:
\begin{equation}
\phi^{ijkl}=\phi_{ijkl}=\sqrt{2}[\phi^{(12)}(\epsilon^{(12)}+\epsilon^{(34)})
+\phi^{(13)}(\epsilon^{(13)}+\epsilon^{(24)})
+\phi^{(14)}(\epsilon^{(14)}+\epsilon^{(23)})]_{ijkl},
\label{eq:scalars}
\end{equation}
where numbers in parentheses correspond to appropriate index pairs
so that {\it e.g.} $\epsilon^{(13)}_{ijkl}=\pm1$ whenever $\{i,j,k,l\}$
corresponds to a permutation of $\{1,2,5,6\}$.  This ansatz is of course
self-dual by construction.  Thus we have reduced the original 70 (real)
scalar degrees of freedom to just three in this specialization.  For this
case, using the definition (\ref{eq:vdv}), we find that the $SU(8)$
connection and scalar kinetic terms become%
\footnote{Note that $SU(8)$ and $SO(8)$ indices are indistinguishable here.
This is a consequence of specializing to a particular gauge choice for the
scalars.}
\begin{equation}
{\cal B}_\mu{}^i{}_j=-2gA_\mu^{ij},\qquad
{\cal A}_\mu{}^{ijkl}=\partial_\mu\phi^{ijkl}.
\end{equation}

When restricted to the abelian $U(1)$ gauge fields, and with the scalar
ansatz (\ref{eq:scalars}), the bosonic lagrangian, (\ref{eq:lag}), may be
rewritten as
\begin{eqnarray}
\label{eq:u1lag}
{\cal L}&=&{1\over2\kappa^2}\sqrt{-g}\Bigl[R-{\textstyle{1\over2}}\left(
(\partial_\mu\phi^{(12)})^2+
(\partial_\mu\phi^{(13)})^2+
(\partial_\mu\phi^{(14)})^2\right)-V\\
&&\qquad-2\left(
e^{-\lambda_1}(F_{\mu\nu}^{(1)})^2+
e^{-\lambda_2}(F_{\mu\nu}^{(2)})^2+
e^{-\lambda_3}(F_{\mu\nu}^{(3)})^2+
e^{-\lambda_4}(F_{\mu\nu}^{(4)})^2
\right)\Bigr],\nonumber
\end{eqnarray}
where the scalar combinations $\{\lambda\}$ are given by
\begin{eqnarray}
\lambda_1&=&-\phi^{(12)}-\phi^{(13)}-\phi^{(14)},\nonumber\\
\lambda_2&=&-\phi^{(12)}+\phi^{(13)}+\phi^{(14)},\nonumber\\
\lambda_3&=&\hphantom{-}\phi^{(12)}-\phi^{(13)}+\phi^{(14)},\nonumber\\
\lambda_4&=&\hphantom{-}\phi^{(12)}+\phi^{(13)}-\phi^{(14)},
\end{eqnarray}
and the scalar potential is
\begin{equation}
V=-4g^2\left(\cosh{\phi^{(12)}}+\cosh{\phi^{(13)}}+\cosh{\phi^{(14)}}\right).
\label{eq:pot}
\end{equation}
Note that the $\{\lambda\}$ are not all independent as
$\lambda_1+\lambda_2+\lambda_3+\lambda_4=0$.  The $U(1)$ gauge fields in
(\ref{eq:u1lag}) are essentially the $SO(8)$ triality rotated combinations
\begin{equation}
\pmatrix{F_{\mu\nu}^{(1)}\cr F_{\mu\nu}^{(2)}\cr
F_{\mu\nu}^{(3)}\cr F_{\mu\nu}^{(4)}}=
{1\over4}\pmatrix{1&\hphantom{-}1&\hphantom{-}1&\hphantom{-}1\cr
1&\hphantom{-}1&-1&-1\cr1&-1&\hphantom{-}1&-1\cr1&-1&-1&\hphantom{-}1}
\pmatrix{F_{\mu\nu}^{12}\cr F_{\mu\nu}^{34}\cr
F_{\mu\nu}^{56}\cr F_{\mu\nu}^{78}}
\equiv{1\over2}\Omega
\pmatrix{F_{\mu\nu}^{12}\cr F_{\mu\nu}^{34}\cr
F_{\mu\nu}^{56}\cr F_{\mu\nu}^{78}}
.
\label{eq:trial}
\end{equation}
For later convenience we have defined the matrix $\Omega$, which satisfies
$\Omega=\Omega^T$ and $\Omega^2=I$.

Several comments are in order here.  The first is that, save for the
potential, (\ref{eq:pot}), and numerical factors in the definition of the
scalars, the truncated bosonic action (\ref{eq:u1lag}) is identical to
that of a closed string compactified on $T^2$ with ``diagonal'' scalars%
\footnote{Note that to make the actual correspondence, two of the gauge
fields (say $F^{(2)}$ and $F^{(4)}$) need to be dualized to provide a
consistent identification with the string dilaton (in this case
$\phi^{(13)}$).  Of course the choice of fields to dualize is only
determined up to string-string-string triality \cite{Duffliurahmfeld1}.}.
While on the one hand this may not be
too surprising, since supersymmetry must necessarily constrain the couplings
between the scalars and vectors, on the other hand it is somewhat remarkable
since the four $U(1)$ fields have rather different interpretations in the
two cases: as the Cartan generators of $SO(8)$ for the gauged supergravity,
and as two Kaluza-Klein and two winding gauge fields for the $T^2$
compactification.  Following up on this correspondence, the second point is
that we have {\it a priori} constrained the three scalars $\phi^{(12)}$,
$\phi^{(13)}$ and $\phi^{(14)}$ to be real.  We believe that
allowing the scalars to be complex in (\ref{eq:scalars}) would in fact
lead to a complete correspondence between (\ref{eq:u1lag}) and the general
$T^2$ compactified effective lagrangian with three complex scalars.
Nevertheless, this brings up the issue that the additional scalar degrees of
freedom do play a role in terms of giving rise to additional conditions on
the bosonic solutions in order to maintain a consistent truncation.
Finally, it is important to realize that when $g\ne0$ the potential
(\ref{eq:pot}) may not be ignored, and fixes the asymptotic scalar values to
vanish, $\phi^{(12)}_\infty=\phi^{(13)}_\infty=\phi^{(14)}_\infty=0$, with
corresponding negative energy density $V_\infty=-12g^2$.

A slight notational complication arises in expressing the supersymmetry
variations (\ref{eq:fsusy}) in terms of explicit field components
(as opposed to fully $SO(8)$ invariant quantities).  As above, we denote
$SO(8)$ index pairs $\{12,34,56,78\}$ by $(\alpha)$ where
$\alpha=1,\ldots,4$, in which case the single $SO(8)$ index $i=1,\ldots,8$
may be replaced by the combination $i_{(\alpha)}$ with the latter $i$ taking
on either 1 or 2 corresponding to the first or the second of the pair
$(\alpha)$.  In this case the gravitino variation may be written as
\begin{eqnarray}
\delta\psi_\mu^{i_{(\alpha)}}&=&
\nabla_\mu\epsilon^{i_{(\alpha)}}-2g\Omega_{\alpha\beta}
A_\mu^{(\beta)}\epsilon_{ij}\epsilon^{j_{(\alpha)}}
+{g\over4\sqrt{2}}\left(e^{\lambda_1/2}+
e^{\lambda_2/2}+ e^{\lambda_3/2}+
e^{\lambda_4/2}\right)\gamma_\mu\epsilon_{i_{(\alpha)}}\nonumber\\
&&\qquad+{1\over2\sqrt{2}}\Omega_{\alpha\beta}
e^{-\lambda_\beta/2}F_{\nu\lambda}^{(\beta)}\gamma^{\nu\lambda}
\gamma_\mu\epsilon^{ij}\epsilon_{j_{(\alpha)}},
\label{eq:gravsusy}
\end{eqnarray}
where the sum over $\beta$ is implied.  For the spin-1/2 fermions and the
Cartan ansatz, we find immediately that $\delta\chi^{ijk}$ vanishes unless
exactly two indices belong in the same pair $(\alpha)$.  Writing the first
two indices as paired, we find
\begin{equation}
\delta\chi^{(\alpha)i_{(\beta)}}=
-{\textstyle{1\over\sqrt{2}}} \gamma^\mu\partial_\mu
\phi^{(\alpha\beta)}\epsilon^{ij}\epsilon_{j_{(\beta)}}
-g\Sigma_{\alpha\beta\gamma}\Omega_{\gamma\delta}e^{\lambda_\delta/2}
\epsilon_{ij}\epsilon^{j_{(\beta)}}
+ \Omega_{\alpha\delta}e^{-\lambda_\delta/2}
F_{\mu\nu}^{(\delta)}\gamma^{\mu\nu}\epsilon^{i_{(\beta)}},
\label{eq:chisusy}
\end{equation}
provided $(\alpha)\ne(\beta)$.  The tensor $\Sigma_{\alpha\beta\gamma}$
selects out a particular $(\gamma)$ depending on $(\alpha\beta)$, and
is defined by
\begin{equation}
\Sigma_{\alpha\beta\gamma}=\cases{
|\epsilon_{\alpha\beta\gamma}|,&for $\alpha,\beta\ne1$\cr
\delta_{\beta\gamma},&for $\alpha=1$\cr
\delta_{\alpha\gamma},&for $\beta=1$.}
\end{equation}

\section{The $a=\protect\sqrt{3}$ black hole in gauged supergravity}

In the previous section we have described a truncation of gauged $N=8$
supergravity to its abelian $U(1)^4$ sector.  The resulting lagrangian and
corresponding fermion supersymmetry may be further simplified by focusing on
single-charge black hole solutions.  In particular, we take
$\phi^{(12)}=\phi^{(13)}=\phi^{(14)}\equiv\phi$ with corresponding
$F_{\mu\nu}^{(1)}\equiv F_{\mu\nu}$ non-vanishing.  The resulting lagrangian
then becomes
\begin{equation}
{\cal L}={1\over2\kappa^2}\sqrt{-g}\left[
R-{3\over2}\partial_\mu\phi\partial^\mu\phi
-2e^{3\phi}F_{\mu\nu}F^{\mu\nu}+12g^2\cosh{\phi}
\right]+\ldots,
\label{eq:sqrt3lag}
\end{equation}
where the dots refer to fields that will be set equal to zero in our
solution.  Note that a scaling $\phi\to\phi/\sqrt{3}$ may be performed to
normalize the scalar kinetic term canonically, which also demonstrates
the correspondence to the conventional $a=\sqrt{3}$ definition, where
$a$ is the scalar-Maxwell parameter appearing in
$e^{a\phi}F_{\mu\nu}F^{\mu\nu}$.

For the single-charge solution, it is straightforward to see that the
supersymmetry transformations, (\ref{eq:gravsusy}) and (\ref{eq:chisusy}),
essentially take on multiple identical copies, one for each
different $(\alpha)$ value.  It is thus sufficient to focus on a single
supersymmetry parameter, {\it e.g.}~$\epsilon^{i_{(1)}}$, instead of the more
general $\epsilon^{i_{(\alpha)}}$.  Note that this also demonstrates how
such a solution can be interpreted in an $N=2$ or $N=4$ context.  The
resulting supersymmetry variations are
\begin{eqnarray}
\half\delta\psi_\mu^i&=&\nabla_\mu\epsilon^i-gA_\mu\epsilon_{ij}
\epsilon^j+{g\over4\sqrt{2}}\left(e^{-3\phi/2}+3e^{\phi/2}
\right)+{1\over4\sqrt{2}}e^{3\phi/2}F_{\nu\lambda}\gamma^{\nu\lambda}
\gamma_\mu\epsilon^{ij}\epsilon_j,\nonumber\\
\delta\chi^i&=&-{1\over\sqrt{2}}\gamma^\mu\partial_\mu\phi\epsilon^{ij}
\epsilon_j
-{g\over2}\left(e^{-3\phi/2}-e^{\phi/2}\right)
\epsilon_{ij}\epsilon^j+{1\over2}e^{3\phi/2}F_{\mu\nu}
\gamma^{\mu\nu}\epsilon^i,
\end{eqnarray}
where we have dropped unnecessary $SO(8)$ index-pair symbols.

Using the well known $a=\sqrt{3}$ black hole solution as a guide, we now
consider a spherically symmetric electric black hole ansatz.
Before describing the black hole in gauged supergravity, we first observe
that when $g=0$ the above lagrangian (\ref{eq:sqrt3lag}) admits an
ordinary $a=\sqrt{3}$ (electric) black hole solution
\cite{Rahmfeld1}:
\begin{equation}
ds^2=-H^{-1/2}dt^2+H^{1/2}(dr^2+r^2d\Omega^2),
\label{eq:sqrt3sol}
\end{equation}
where $H=1+Q/r$, and with the scalar and gauge field given (in the above
normalization) by
\begin{equation}
e^{2(\phi-\phi_\infty)}=H,\qquad
A_0={\eta\over2\sqrt{2}}e^{-3\phi_\infty/2}H^{-1}
\label{eq:hfa}
\end{equation}
($\eta=\pm1$ sets the actual sign of the charge).  Furthermore, we recall
that it was demonstrated in \cite{Romans,Behrndt1,Behrndt2}
that extreme
black hole solutions generally have simple extensions to the case of
non-zero gauging, and in fact generally retain their supersymmetry
properties.  Following \cite{Behrndt1,Behrndt2}, we thus take the
metric ansatz
\begin{equation}
ds^2=-e^{2A}fdt^2+e^{-2A}({dr^2\over f}+r^2d\Omega^2),
\label{eq:adsmet}
\end{equation}
corresponding to an $AdS$ generalization of (\ref{eq:sqrt3sol}).  Note in
particular that the vacuum $AdS$ solution is given by the choice $f=1+2g^2r^2$
with $A=0$.

With the metric (\ref{eq:adsmet}) the gauge equation of
motion becomes $\partial_r(e^{-2A+3\phi}r^2\partial_rA_0)=0$ and
is unaffected by both $g$ and the function $f$.  This suggests that the
harmonic function ansatz, (\ref{eq:hfa}), simply carries over to the
$g\ne0$ case, with the only additional constraint that $\phi_\infty=0$.
Turning to the supersymmetry variation, $\delta\chi^i$,
%
%
we find, using (\ref{eq:hfa}), that
\begin{equation}
\delta\chi^i=-{1\over2\sqrt{2}}{\partial_r H\over H}\epsilon^{ij}
\gamma^r\left[
\epsilon_j+f^{-1/2}(\eta\gamma_{\overline{0}}\epsilon_{jk}\epsilon^k
+\sqrt{2}grH^{1/2}\gamma_r\epsilon^j)\right],
\end{equation}
so that the natural half-supersymmetry projection is given by
\begin{equation}
P_\eta^{ij}=\half[\delta^{ij}+f^{-1/2}(\eta\gamma_{\overline{0}}\epsilon_{ij}
+\sqrt{2}grH^{1/2}\gamma_{\overline{r}}\delta^{ij})]
\label{eq:halfsusy}
\end{equation}
(acting on real spinors $\epsilon^i=\epsilon_i$) provided $f=1+2g^2r^2H$.
Using this expression for $f$, it is now straightforward to check that all
bosonic equations of motion arising from (\ref{eq:sqrt3lag}) are satisfied.
To summarize, the single-charge black hole solution in gauged supergravity
is given by
\begin{eqnarray}
&&ds^2=-H^{-1/2}fdt^2+H^{1/2}({dr^2\over f}+r^2d\Omega^2),\nonumber\\
&&e^{2\phi}=H,\qquad
A_0={\eta\over2\sqrt{2}}H^{-1},
\label{eq:1charge}
\end{eqnarray}
where
\begin{equation}
H=1+{Q\over r}, \qquad f=1+2g^2r^2H.
\end{equation}

We now turn to an examination of the supersymmetry properties of this
solution.  In addition to $\delta\chi^i$ given above:
\begin{equation}
\delta\chi^i={Q\over\sqrt{2}r^2}H^{-1}\epsilon^{ij}\gamma^rP_\eta\epsilon_j,
\end{equation}
the gravitino variations in the background (\ref{eq:1charge}) are given by
\begin{eqnarray}
\half\delta\psi_0^i&=&\partial_0\epsilon^i+{g\eta\over2\sqrt{2}}\epsilon_{ij}
\epsilon^j+{g\over\sqrt{2}}H^{-1}f^{1/2}(1+H)\gamma_{\overline{0}}
P_\eta\epsilon_i+{Q\over4r^2}H^{-3/2}f\gamma_{\overline{0r}}P_\eta
\epsilon^i,\nonumber\\
\half\delta\psi_r^i&=&(\partial_r-{Q\over8r^2}H^{-1})\epsilon^i
+{g\over2\sqrt{2}}H^{-1/2}f^{-1/2}(1+H)\gamma_{\overline{r}}\epsilon_i
+{Q\over4r^2}H^{-1}P_\eta\epsilon^i,\nonumber\\
\half\delta\psi_\theta^i&=&\partial_\theta\epsilon^i-{\eta\over2}
\gamma_{\overline{0\theta r}}\epsilon_{ij}\epsilon_j
+(1-{Q\over4r}H^{-1})f^{1/2}\gamma_{\overline{\theta r}}P_\eta\epsilon^i,
\nonumber\\
\half\delta\psi_\phi^i&=&\partial_\phi\epsilon^i-{\eta\over2}\sin\theta
\gamma_{\overline{0\phi r}}\epsilon_{ij}\epsilon_j+{1\over2}
\cos\theta\gamma_{\overline{\phi\theta}}\epsilon^i
+(1-{Q\over4r}H^{-1})f^{1/2}\sin\theta \gamma_{\overline{\phi r}}
P_\eta\epsilon^i.
\label{eq:1gravar}
\end{eqnarray}
For Killing spinors, $P_\eta\epsilon_i=0$, we may follow the construction
of \cite{Romans,Behrndt1,Behrndt2}, to obtain
\begin{equation}
\epsilon=
e^{-{g\eta t\over2\sqrt{2}}\epsilon_{ij}}H^{-1/8}
\left[\sqrt{f^{1/2}+1}-\sqrt{f^{1/2}-1} \gamma_{\overline{r}}\right]
e^{-{1\over2}\gamma_{\overline{\theta r}}}
e^{-{1\over2}\gamma_{\overline{\phi\theta}}}
(1-\eta\gamma_{\overline{0}}\epsilon^{ij})\epsilon_0,
\label{eq:1kil}
\end{equation}
so that this solution in fact preserves exactly half of the supersymmetries.
Note that substituting $\epsilon_0={1\over\sqrt{2}}(1-\gamma_{\overline{r}})
\tilde\epsilon_0$ and using the identity $\sqrt{f^{1/2}+1}\pm\sqrt{f^{1/2}-1}
=\sqrt{2(f^{1/2}\pm(f-1)^{1/2})}$ indicates that (\ref{eq:1kil}) may
be rewritten equivalently as
\begin{equation}
\epsilon=
e^{-{g\eta t\over2\sqrt{2}}\epsilon_{ij}}H^{-1/8}
\left[\sqrt{f^{1/2}+(f-1)^{1/2}}-\eta\sqrt{f^{1/2}-(f-1)^{1/2}}
\gamma_{\overline{0}}\epsilon^{ij}\right]
e^{{\eta\over2} \gamma_{\overline{0\theta r}}\epsilon_{ij}}
e^{-{1\over2}\gamma_{\overline{\phi\theta}}}
(1-\gamma_{\overline{r}})\tilde\epsilon_0,
\end{equation}
which is the form that appears in \cite{Romans}.  Although we have
focused on $N=8$ in the present case, this Killing spinor construction
is general, and also applies in the $N=2$ and $N=4$ truncations of the
full $N=8$ theory.

When written in the form (\ref{eq:1kil}), the above Killing spinors resemble
a supersymmetry projected version of the corresponding Killing spinors in
pure Anti-de Sitter space%
\footnote{
Some of the difference in the $t$-dependent terms may be eliminated by a
suitable gauge transformation.  This issue also arises in the next section
when considering the multiple charge black hole.}:
\begin{equation}
\epsilon({AdS})=
\left[\sqrt{f^{1/2}+1}-\sqrt{f^{1/2}-1} \gamma_{\overline{r}}\right]
e^{-{1\over2} \gamma_{\overline{\theta r}}}
e^{-{1\over2} \gamma_{\overline{\phi\theta}}}
e^{-{gt\over\sqrt{2}}\gamma_{\overline{0}}}
\epsilon_0.
\end{equation}
Additionally, the Killing spinors may be contrasted with those arising
in the ungauged theory.  Taking $g\to0$, we find
\begin{equation}
\epsilon(g=0)= H^{-1/8}
e^{-{1\over2} \gamma_{\overline{\theta r}}}
e^{-{1\over2} \gamma_{\overline{\phi\theta}}}
(1-\eta\gamma_{\overline{0}}\epsilon^{ij})
\epsilon_0,
\label{eq:rigideps}
\end{equation}
which satisfies the well known Killing spinor condition
$P_\eta^0\epsilon(g=0)\equiv\half(1+\eta\gamma_{\overline{0}}
\epsilon^{ij})\epsilon(g=0)=0$.  An added consequence of this simple
structure in the $g\to0$ case is that the fermion zero modes are easily
constructed solely by changing the sign of the projection in
(\ref{eq:rigideps}).  Such zero modes are immediately orthogonal to the
Killing spinors and furthermore satisfy the supergauge condition
$\gamma^\mu\delta\psi_\mu^i=0$ \cite{Gibbons,Aichelburg}.  Unfortunately
this situation is not as clear when $g\ne0$; this is mainly due to
complications arising from the nature of the projection (\ref{eq:halfsusy})
in the gauged supergravity.  For this reason, although it is manifest that
this black hole preserves exactly half of the supersymmetries, its
supermultiplet structure arising from the fermion zero mode construction
\cite{Aichelburg, Duffliurahmfeld2} is less well understood.  On the
other hand, the fact that it preserves half the supersymmetry
presumably means that it belongs to the short maximum spin 2
supermultiplet.

\section{Multiple charge black holes}

Returning to the complete simplified $N=8$ lagrangian (\ref{eq:u1lag}) we
note that, in the absence of the scalar potential, this admits well known
supersymmetric black hole solutions with up to four charges.  Since the
single charge solution has a straightforward generalization for $g\ne0$, as
shown above, one may wonder whether this is also true for the four charge
solution.  A careful examination of the equations of motion
arising from (\ref{eq:u1lag}) shows that this is in fact the case.  In
contrast to the usual form of the action arising from $T^2$ compactification
of the closed string (which may be written in either $S$, $T$ or $U$ form
\cite{Duffliurahmfeld1}), in which a dilaton scalar is singled out, the
Lagrangian
(\ref{eq:u1lag}) treats all three scalars and four gauge fields
symmetrically.  In practice, this indicates that we are interested in a
four electric charge black hole solution.  We find
\begin{eqnarray}
&&ds^2=-(H_1H_2H_3H_4)^{-1/2}fdt^2+(H_1H_2H_3H_4)^{1/2}
({dr^2\over f}+r^2d\Omega^2),\nonumber\\
&&e^{2\phi^{(12)}}={H_1H_2\over H_3H_4},\qquad
e^{2\phi^{(13)}}={H_1H_3\over H_2H_4},\qquad
e^{2\phi^{(14)}}={H_1H_4\over H_2H_3},\nonumber\\
&&
A_0^{(\alpha)}={\eta_\alpha\over2\sqrt{2}}H_\alpha^{-1},
\label{eq:4charge}
\end{eqnarray}
where
\begin{equation}
H_\alpha=1+{Q_\alpha\over r}, \qquad f=1+2g^2r^2(H_1H_2H_3H_4).
\label{eq:4harm}
\end{equation}

{}From the decomposition of the $N=8$ spinor parameter $\epsilon_i$ into the
four $\epsilon_{i_{(\alpha)}}$ where $\alpha=1,\ldots,4$, we see that the
$N=8$ supersymmetry variations, (\ref{eq:gravsusy}) and
(\ref{eq:chisusy}), break up into four sets, involving separate $\pm$ signs
in the combination of the field strengths.  Focusing on a single set of
variations, we find
\begin{eqnarray}
\half\delta\psi_\mu^{i_{(1)}}&=&\nabla_\mu\epsilon^{i_{(1)}}
-g(A_\mu^{(1)}+A_\mu^{(2)}+A_\mu^{(3)}+A_\mu^{(4)})\epsilon_{ij}
\epsilon^{j_{(1)}}\nonumber\\
&&+{g\over4\sqrt{2}}\left(e^{\lambda_1/2}+
e^{\lambda_2/2}+ e^{\lambda_3/2}+
e^{\lambda_4/2}\right)\gamma_\mu\epsilon_{i_{(1)}}\nonumber\\
&&+{1\over4\sqrt{2}}\left(e^{-\lambda_1/2}F_{\nu\lambda}^{(1)}+
e^{-\lambda_2/2}F_{\nu\lambda}^{(2)}+
e^{-\lambda_3/2}F_{\nu\lambda}^{(3)}+
e^{-\lambda_4/2}F_{\nu\lambda}^{(4)}\right)\gamma^{\nu\lambda}
\gamma_\mu\epsilon^{ij}\epsilon_{j_{(1)}},\nonumber\\
\delta(2\chi^{(3)i_{(1)}})&=&-\sqrt{2}
\gamma^\mu\partial_\mu\phi^{(13)}\epsilon^{ij}
\epsilon_{j_{(1)}}-g\left((e^{\lambda_1/2}+e^{\lambda_3/2})
-(e^{\lambda_2/2}+e^{\lambda_4/2})\right)
\epsilon_{ij}\epsilon^{j_{(1)}}\nonumber\\
&&+ \left( (e^{-\lambda_1/2}F_{\mu\nu}^{(1)}+
e^{-\lambda_3/2}F_{\mu\nu}^{(3)})
-( e^{-\lambda_2/2}F_{\mu\nu}^{(2)}+
e^{-\lambda_4/2}F_{\mu\nu}^{(4)})\right)
\gamma^{\mu\nu}\epsilon^{i_{(1)}},\nonumber
\end{eqnarray}
\vspace{-24pt}
\begin{eqnarray}
\delta(\chi^{(2)i_{(1)}}+\chi^{(4)i_{(1)}})&=&
-{\textstyle{1\over\sqrt{2}}}
\gamma^\mu\partial_\mu(\phi^{(12)}+\phi^{(14)})\epsilon^{ij}
\epsilon_{j_{(1)}}-g\left(e^{\lambda_1/2}-e^{\lambda_3/2}
\right)\epsilon_{ij}\epsilon^{j_{(1)}}\nonumber\\
&&+ \left( e^{-\lambda_1/2}F_{\mu\nu}^{(1)}-
e^{-\lambda_3/2}F_{\mu\nu}^{(3)}\right)
\gamma^{\mu\nu}\epsilon^{i_{(1)}},\nonumber\\
\delta(\chi^{(2)i_{(1)}}-\chi^{(4)i_{(1)}})&=&
-{\textstyle{1\over\sqrt{2}}}
\gamma^\mu\partial_\mu(\phi^{(12)}-\phi^{(14)})\epsilon^{ij}
\epsilon_{j_{(1)}}-g\left(e^{\lambda_2/2}-e^{\lambda_4/2}
\right)\epsilon_{ij}\epsilon^{j_{(1)}}\nonumber\\
&&+ \left( e^{-\lambda_2/2}F_{\mu\nu}^{(2)}-
e^{-\lambda_4/2}F_{\mu\nu}^{(4)}\right)
\gamma^{\mu\nu}\epsilon^{i_{(1)}}.
\end{eqnarray}
The choice of the particular linear combinations of the spin-1/2 supersymmetry
variations used above is motivated by the correspondence to the $N=4$ theory
arising from $T^2$ compactification:
\begin{eqnarray}
2\chi^{(3)}&\longleftrightarrow&\lambda\hbox{ (dilatino)}\nonumber\\
\chi^{(2)}\pm\chi^{(4)}&\longleftrightarrow&\tilde\chi^{1,2}\hbox{ (gauginos)}
\nonumber\\
\phi^{(13)}&\longleftrightarrow&\eta\hbox{ (dilaton)}\nonumber\\
\phi^{(12)},\phi^{(14)}&\longleftrightarrow&\rho,\sigma\hbox{ (internal $T^2$
metric)}.
\end{eqnarray}

For the four-charge solution (\ref{eq:4charge}), we find that preserving
supersymmetry (in the $\epsilon_{i_{(1)}}$ sector) demands a particular
choice of signs for the charges, $\eta_1=\eta_2=\eta_3=\eta_4$ ($=\eta$), in
which case we find
\begin{eqnarray}
\delta(2\chi^{(3)i_{(1)}})&=&-\sqrt{2}\epsilon^{ij}\gamma^r\partial_r
\log{H_1H_3\over H_2H_4}P_\eta\epsilon_{j_{(1)}},\nonumber\\
\delta(\chi^{(2)i_{(1)}}+\chi^{(4)i_{(1)}})
&=&-\sqrt{2}\epsilon^{ij}\gamma^r\partial_r
\log{H_1\over H_3}P_\eta\epsilon_{j_{(1)}},\nonumber\\
\delta(\chi^{(2)i_{(1)}}-\chi^{(4)i_{(1)}})
&=&-\sqrt{2}\epsilon^{ij}\gamma^r\partial_r
\log{H_2\over H_4}P_\eta\epsilon_{j_{(1)}},
\label{eq:4cchisusy}
\end{eqnarray}
where
\begin{equation}
P_\eta^{ij}=\half[\delta^{ij}+f^{-1/2}(\eta\gamma_{\overline{0}}\epsilon_{ij}
+\sqrt{2}gr{\cal H}^{1/2}\gamma_{\overline{r}}\delta^{ij})],
\label{eq:4cproj}
\end{equation}
with ${\cal H}=H_1H_2H_3H_4$.
For the four charge solution the gravitino variations are
\begin{eqnarray}
\half\delta\psi_0^{i_{(1)}}&=&\partial_0\epsilon^{i_{(1)}}
-{g\eta\over\sqrt{2}} \epsilon_{ij}
\epsilon^{j_{(1)}}+{g\over\sqrt{2}}f^{1/2}(2+r\partial_r\log{\cal H})
\gamma_{\overline{0}}
P_\eta\epsilon_{i_{(1)}}-{1\over4}{\cal H}^{-1/2}f\partial_r\log{\cal H}
\gamma_{\overline{0r}}P_\eta \epsilon^{i_{(1)}},\kern-8pt\nonumber\\
\half\delta\psi_r^{i_{(1)}}&=&(\partial_r+{1\over8}\partial_r\log{\cal H})
\epsilon^{i_{(1)}}
+{g\over2\sqrt{2}}{\cal H}^{1/2}f^{-1/2}(2+r\partial_r\log{\cal H})
\gamma_{\overline{r}}\epsilon_{i_{(1)}}
-{1\over4}\partial_r\log{\cal H}P_\eta\epsilon^{i_{(1)}},\nonumber\\
\half\delta\psi_\theta^{i_{(1)}}&=&\partial_\theta\epsilon^{i_{(1)}}
-{\eta\over2} \gamma_{\overline{0\theta r}}\epsilon_{ij}\epsilon_{j_{(1)}}
+{1\over4}f^{1/2}(4+r\partial_r\log{\cal H})\gamma_{\overline{\theta r}}
P_\eta\epsilon^{i_{(1)}},\\
\half\delta\psi_\phi^{i_{(1)}}&=&\partial_\phi\epsilon^{i_{(1)}}
-{\eta\over2}\sin\theta
\gamma_{\overline{0\phi r}}\epsilon_{ij}\epsilon_{j_{(1)}}+{1\over2}
\cos\theta\gamma_{\overline{\phi\theta}}\epsilon^{i_{(1)}}
+{1\over4}f^{1/2}(4+r\partial_r\log{\cal H})\sin\theta
\gamma_{\overline{\phi r}}
P_\eta\epsilon^{i_{(1)}}.\nonumber
\end{eqnarray}
In certain cases we have used the identity
$r\partial_rH_\alpha=1-H_\alpha$ [which holds for $H_\alpha$ given in
(\ref{eq:4harm})] when combining some of the factors
in $\half\delta\psi_\mu^{i_{(1)}}$ to form the half-supersymmetry
projection terms.

We see that the Killing spinor equations, $\half\delta\psi_\mu^{i_{(1)}}=0$
with $P_\eta\epsilon_{i_{(1)}}=0$, are practically identical with those that
arise from the single charge case, (\ref{eq:1gravar}).  Thus the
Killing spinors are similar to those of (\ref{eq:1kil}) and have the form
\begin{equation}
\epsilon^{(1)}=
e^{{g\eta t\over\sqrt{2}}\epsilon_{ij}}{\cal H}^{-1/8}
\left[\sqrt{f^{1/2}+1}-\sqrt{f^{1/2}-1} \gamma_{\overline{r}}\right]
e^{{\eta\over2} \gamma_{\overline{0\theta r}}\epsilon^{ij}}
e^{-{1\over2}\gamma_{\overline{\phi\theta}}}
(1-\eta\gamma_{\overline{0}}\epsilon^{ij})\epsilon_0^{(1)}.
\end{equation}
Until now we have only considered the first out of four sets of $N=8$
supersymmetries, namely those parametrized by $\epsilon_{i_{(1)}}$.  Naturally
the form of the other three sets of variations are constrained by $N=8$
supersymmetry, and differ only by the relative choices of signs between
the four Cartan gauge fields.  Preservation of (half) supersymmetry in
each of the four sectors demands the following sign choices:
\begin{eqnarray}
&&1:\quad\eta_1=\hphantom{-}\eta_2=\hphantom{-}\eta_3=\hphantom{-}\eta_4
\nonumber\\
&&2:\quad\eta_1=\hphantom{-}\eta_2=-\eta_3=-\eta_4\nonumber\\
&&3:\quad\eta_1=-\eta_2=\hphantom{-}\eta_3=-\eta_4\nonumber\\
&&4:\quad\eta_1=-\eta_2=-\eta_3=\hphantom{-}\eta_4.
\end{eqnarray}
It is not coincidental that these signs match those of $\Omega$ defined in
(\ref{eq:trial}).
Because of the necessary difference in signs above, this indicates that
in general, when all four charges are active, supersymmetry cannot be
partially preserved in all sectors simultaneously.  For one through four
active charges, we find overall that 1/2, 1/4, 1/8 and 1/8 of the $N=8$
supersymmetry can be preserved, in complete agreement with standard results
\cite{Lupope,Khuriortin}. When all charges are equal, the solution may
be obtained from a single scalar, single Maxwell field truncation with
scalar-Maxwell parameter $a=\sqrt{3},1,1/\sqrt{3},0$ just as in the
case of non-gauged supergravity
\cite{Rahmfeld1,Rahmfeld2,Rahmfeld3,Duffliurahmfeld1}.
Of course, one can also choose the
charges so that fewer or even no supersymmetries are preserved,
even though the black holes are still extremal \cite{Rahmfeld1}.

In the case of ungauged supergravity, on the basis of these mass and charge
assignments, it was further suggested
\cite{Rahmfeld1,Rahmfeld2,Rahmfeld3,Duffliurahmfeld1} that we interpret
these four values of $a$ as 1-, 2-, 3- and $4$-particle {\it bound states}
with zero binding
energy. For example, the Reissner-Nordstrom ($a=0$) black hole
combines four ($a=\sqrt{3}$) black holes: an electric Kaluza-Klein
black hole, a magnetic
Kaluza-Klein black hole, an electric winding black hole and a magnetic
winding black hole. This zero-binding-energy bound-state conjecture can,
in fact, be verified in the classical black hole picture
by finding explicit $4$-centered black hole solutions which coincide
with the $a=\sqrt{3},1,1/\sqrt{3},0$ solutions as we bring $1,2,3,4$
centers together and take the remaining $3,2,1,0$ centers out to infinity
\cite{Rahmfeld2}. Such a construction is possible because of the
appearance of four independent harmonic functions
\cite{Cvetictseytlin}. Moreover, this provides a novel realization of the {\it
no-force} condition in that the charge carried by each black hole corresponds
to a different $U(1)$. Thus the gravitational attraction cannot be cancelled
by an electromagnetic repulsion but rather by a subtle repulsion due to scalar
exchange. This phenomenon was also observed in \cite{Kalloshlinde}.  In the
above, for purposes of illustration, the special case has been chosen where
all non-zero charges are equal to unity but it is easily generalized to the
case of different electric charges $Q_1,P_2,Q_3,P_4$ where the interpretation
is that of a $(Q_1+P_2+Q_3+P_4)$-particle bound state with zero binding energy
\cite{Cvetic}.

It is tempting to generalize this bound state picture to the black holes
of the gauged supergravity discussed in the present paper.  One
interesting difference from the non-gauged supergravity case,
however, is that we find no static multi-center solutions.  While the
present solution is again based on four harmonic functions,
(\ref{eq:4harm}), they are however not strictly independent as they must all
share the same center%
\footnote{Examination of the equations of motion indicates that the
obstruction to finding multi-center solutions arises not from the gauge
equation, but rather from the scalar and Einstein equations.}.
This is to be expected on physical grounds: the
presence of a negative cosmological constant ensures that only single center
solutions will be static.

\section{The non-extremal solution}

While we are mainly interested in the properties of supersymmetric black
holes, we note that there is a straightforward generalization of the above
solutions to the non-extremal case.  In the ungauged case, the extremal
solution can be ``blackened'' by the incorporation of a universal function
$f=1-k/r$ modifying the standard $p$-brane metric of the form
\cite{dufflupope}
\begin{equation}
ds^2=-e^{2A}fdt^2+e^{2B}({dr^2\over f}+r^2d\Omega^2).
\label{eq:nonextr}
\end{equation}
Subsequently, it was shown in \cite{Behrndt2} that this prescription
generalizes in the straightforward manner to $AdS$ black hole solutions as
well.  In particular, note that the four-charge black hole metric of
(\ref{eq:4charge}) has the identical form as the non-extremal metric
(\ref{eq:nonextr}) and hence appears compatible with the ``blackening''
procedure.

As the manipulations of the equations of motion arising from (\ref{eq:u1lag})
are not particularly illuminating, we only present the result here.  The
essential feature of the non-extremal black hole solution is a merging of
the $AdS$ function $f=1+2g^2r^2(H_1H_2H_3H_4)$ of (\ref{eq:4harm}) with the
non-extremal function $f=1-k/r$ to arrive at
\begin{equation}
f=1-{k\over r}+2g^2r^2(H_1H_2H_3H_4).
\end{equation}
In addition, there is a charge rescaling so that the physical electric
charges are no longer related to the mass.  Introducing $\mu_\alpha$
($\alpha=1,\ldots,4$) to parametrize the four charges, we may write
\begin{equation}
H_\alpha=1+{k\sinh^2\mu_\alpha\over r},\qquad
A_0^{(\alpha)}={\eta_\alpha\over2\sqrt{2}}\coth\mu_\alpha H_\alpha^{-1},
\end{equation}
so that
\begin{equation}
F_{0r}^{(\alpha)}=-{\eta_\alpha\over2\sqrt{2}}H_\alpha^{-2}
{k\cosh\mu_\alpha\sinh\mu_\alpha\over r^2}.
\end{equation}
Note that the extremal limit is approached by letting $k\to0$ and
$\mu_\alpha\to\infty$  with $Q_\alpha\equiv k\sinh^2\mu_\alpha$ fixed.

Spacetime properties of the above black holes depend on the number of active
charges ($n=0,\ldots,4$).  For $n=0$ the solution reduces to the
Schwarzschild-anti-de Sitter black hole with a single horizon protecting the
singularity at $r=0$.  On the other hand, the BPS solutions ($k=0$) for
$n=1,2,3$ all have singular horizons at $r=0$ (with zero area) as
appropriate for extremal black holes.  Somewhat surprisingly, though, the
four-charge BPS black hole has no horizon, and so strictly speaking is a
naked singularity%
\footnote{For four identical charges, the solution (\ref{eq:4charge})
reduces to the Reissner-Nordstrom-anti-de Sitter black hole whose
properties were studied in \cite{Romans}.  That Einstein-Maxwell theory
with a cosmological constant is a consistent truncation of $D=11$
supergravity, and hence that $D=11$ supergravity has the
Reissner-Nordstrom-anti-de Sitter black hole as a
solution, has been known for some time \cite{Duffpope,Popetrunc}.}.
In all cases, the existence of a regular horizon demands $k>k_{min}$ where
$k_{min}$ is a function of the active charges.
This is in fact similar to the five dimensional case considered in
\cite{Behrndt1,Behrndt2}.

\section{Magnetic black holes}

We have seen that the $N=8$ gauged supergravity naturally admits a
four-electric-charge black hole solution.  In fact it turns out that
this solution is easily generalized to give magnetically charged black
holes; although the full theory involves non-abelian $SO(8)$ gauge
fields, the $U(1)^4$ truncation of (\ref{eq:u1lag}) gives rise to bosonic
equations of motion that are symmetric under the electric-magnetic duality
\begin{equation}
F^{(\alpha)}\to e^{-\lambda_\alpha}*F^{(\alpha)},
\qquad\lambda_\alpha\to-\lambda_\alpha.
\end{equation}
The resulting four magnetic charge solution has the form
\begin{eqnarray}
&&ds^2=-(H_1H_2H_3H_4)^{-1/2}fdt^2+(H_1H_2H_3H_4)^{1/2}
({dr^2\over f}+r^2d\Omega^2),\nonumber\\
&&e^{2\phi^{(12)}}={H_3H_4\over H_1H_2},\qquad
e^{2\phi^{(13)}}={H_2H_4\over H_1H_3},\qquad
e^{2\phi^{(14)}}={H_2H_3\over H_1H_4},\nonumber\\
&&H_\alpha=1+{k\sinh^2\mu_\alpha\over r},
\qquad f=1-{k\over r}+2g^2r^2(H_1H_2H_3H_4),\nonumber\\
&&F_{\theta\phi}^{(\alpha)}={\eta_\alpha\over2\sqrt{2}}
k\cosh\mu_\alpha\sinh\mu_\alpha\sin\theta.
\la{magnetic}
\end{eqnarray}

While the extremal limit is once again reached by taking $k\to0$ and
$\mu_\alpha\to\infty$ with $P_\alpha\equiv k\sinh^2\mu_\alpha$ fixed, the
resulting extremal black hole is in fact not supersymmetric whenever
$g\ne0$! In the case of the magnetic Reissner-Nordstrom black hole,
this phenomenon was previously found in \cite{Romans}.  (Note,
however, that
it is possible to obtain magnetic black holes that do preserve some
supersymmetry if one allows for event horizons with non-spherical
topologies \cite{Caldarelli})
To see that (\ref{magnetic}) admits no Killing spinors,
we note that while the scalar potential (\ref{eq:pot}) is symmetric under
$\phi^{(\alpha\beta)}\to-\phi^{(\alpha\beta)}$, the scalar related
terms in the supersymmetry variations (\ref{eq:gravsusy}) and
(\ref{eq:chisusy}) are not.  In particular, focusing on
$\delta\chi$, we find for example
\begin{eqnarray}
\delta(2\chi^{(3)i_{(1)}})&=&{1\over\sqrt{2}}\epsilon^{ij}\gamma^r
\Biggl\{\partial_r\log{H_1H_3\over H_2H_4} [\delta^{jk}-i\eta f^{-1/2}
\gamma_{\overline{0}}\gamma^5\epsilon_{jk}]\nonumber\\
&&\qquad\qquad+\partial_r((H_1+H_3)-(H_2+H_4))
[\sqrt{2}grf^{-1/2} \gamma_{\overline{r}}\delta^{jk}] \Biggr\}
\epsilon_{k_{(1)}}
\label{eq:magsusy}
\end{eqnarray}
(where $\eta\equiv\eta_1=\eta_2=\eta_3=\eta_4$),
indicating explicitly that the $g$-dependent term on the last line has a
different structure than the others.  This is in contrast with
(\ref{eq:4cchisusy})
for the electric solution where all terms combine to give the projection
operator (\ref{eq:4cproj}).  Additionally, note that the matrices
$[i\gamma_{\overline{0}}\gamma^5\epsilon_{ij}]$ and
$[\gamma_{\overline{r}}\delta^{ij}]$ now commute, while previously, for the
electric black hole, they had anticommuted in the absence of $\gamma^5$.

For both of the above reasons, we see that whenever $g\ne0$ none of the
supersymmetry variations vanish, and hence the magnetic solution is
non-BPS (regardless of the choice of signs of the magnetic charges).  In the
$g\to0$ limit, on the other hand, the last line of (\ref{eq:magsusy}) drops
out, and we are left with
\begin{equation}
\delta(2\chi^{3i_{(1)}})=\sqrt{2}\epsilon^{ij}\gamma^r\partial_r
\log{H_1H_3\over H_2H_4}\tilde P_\eta\epsilon_{j_{(1)}},
\end{equation}
where $\tilde P_\eta^{ij}={1\over2}[\delta^{ij}-i\eta\gamma_{\overline{0}}
\gamma^5\epsilon_{ij}]$ is the projection appropriate to a magnetically
charged solution.  Thus, in the absence of gauging, there is a direct
correspondence between the supersymmetry properties of the electric and the
magnetic black holes.  However the gauged supergravity theory (in the
abelian truncation) is apparently no longer invariant under
electric-magnetic duality.

\section{Kaluza-Klein states as black holes}

For {\it ungauged} $N=8$ supergravity, the supersymmetry algebra
admits 4 central charges $Z_{1},Z_{2},Z_{3},Z_{4}$. States fall
into 5 categories according as they are annihilated by $4 \geq q \geq
0$ supersymmetry generators. $q$ also counts the number of $Z$'s that
obey the bound $M=Z_{max}$. Non-rotating black holes (in the
sense of vanishing bosonic Kerr angular momentum $L$) belong to
superspin $L=0$ supermultiplets \cite{Rahmfeld3,Duffliurahmfeld1}. Starting
with a spin $J=0$ member, the rest of the black hole multiplet may
then be filled out using the fermionic zero-modes
\cite{Aichelburg,Duffliurahmfeld2}. The spin will run from $J=0$ up
to $J=(8-q)/2$. For {\it gauged} supergravity, the algebra is different
with no central charges but the same multiplet shortening phenomenon
still occurs \cite{Freedmannicolai,Duffnilssonpope}. So we can be
confident that the above black holes preserving $4,2,1,0$ supersymmetries will
belong to supermultiplets with maximum spins $2,3,7/2,4$
\footnote{The maximum spin $5/2$ solutions are (mysteriously?) absent
just as for ungauged supergravity.}. Unfortunately, as far as we know,
the analogue of the $M=Z_{max}$ condition has never been been spelled out
in the literature. It is presumably some relation between the $AdS$ quantum
numbers $(E_{0},s)$ and the $SO(8)$ Casimirs.

It seems entirely consistent, therefore, to identify a subset of the maximum
spin 2 black hole supermultiplets with the $S^{7}$ Kaluza-Klein
spectrum, in analogy with the black hole Kaluza-Klein correspondence
of ungauged supergravity \cite{Rahmfeld1,Kaluza}. The subset in
question will correspond to electric black holes whose mass is quantized
in units of the inverse $S^7$ radius. However, this raises the
puzzle of how the black holes carrying
only $U(1)$ charges can be identified with the Kaluza-Klein particles
carrying non-trivial $SO(8)$ representations. Although we have not
demonstrated this explicitly, it seems reasonable to suppose that it
is the fermion zero modes that provide the non-trivial $SO(8)$ quantum
numbers just as they provide the non-trivial spin. The fact that
these nonabelian charges arise from
{\it fermionic hair} also
nicely circumvents the usual no-hair theorems of classical
relativity. In this connection, it would be interesting to repeat the
gyromagnetic ratio calculations of \cite{Duffliurahmfeld2} and verify
that the fermionic hair again yields a gyromagnetic ratio equal to 1, as
demanded by Kaluza-Klein reasoning.

It is furthermore tempting, in analogy with the ungauged case, to
identify the 2, 3 and 4 charge solutions as 2, 3 and 4-particle
bound states of the singly charged solution
\cite{Duffliurahmfeld1,Rahmfeld2}. However, although the quantum number
assignments are consistent with this, we do not have multi-center
solutions in the $AdS$ case. Such a bound state interpretation would,
of course, lead to states of arbitrarily high spin.

Another difference between the $S^{7}$ and the $T^{7}$
compactifications is that the $g\rightarrow 0$ limit of the gauged
supergravity does not directly coincide with the massless sector of
the $T^{7}$ compactification. They differ by various dualizations. Thus
it was possible, for example, to find 4-charge solutions with all
charges electric as opposed to the 2-electric and 2-magnetic charges
of the ungauged theory. Moreover, whereas 2 charges are Kaluza-Klein
modes and 2 are winding modes in the Type $IIA$ string theory
context, there is no T or U-duality associated with the $S^7$
compactification.

One might also generalize the purely electric and purely
magnetic solutions of this paper to dyonic black hole solutions of
gauged $N=8$ supergravity.  Neither magnetic nor dyonic black holes
have any Kaluza-Klein interpretation and do not appear in the spectrum
of the $S^{7}$ compactification of $D=11$ supergravity. It would be
interesting to provide their M-theory interpretation and to determine their
role in the $AdS$/CFT correspondence. Might they be related to a
Goddard-Nuyts-Olive \cite{Goddardnuytsolive} non-abelian duality, for example?
Since, in the $g \rightarrow 0$
limit, we recover the black holes which were previously identified
with the $T^7$ spectrum, moreover, it seems possible that M-theory can
interpolate between the two topologies. Perhaps the $D=11$
supermembrane, which interpolates between $AdS^{7}\times S^{7}$ and
flat spacetime \cite{GT,DGT,GHT}, plays an important part in this.

\bigskip
\noindent {\bf Acknowledgement}

We have enjoyed useful conversations with Joachim Rahmfeld.

\newpage


\begin{thebibliography}{99}

\bibitem{Maldacena}
J. Maldacena,
{\sl The large $N$ limit of superconformal field theories and
supergravity}, Adv. Theor. Math. Phys. 2 (1998) 231.

\bibitem{Gubser}
S. S. Gubser, I. R. Klebanov and A. M. Polyakov,
{\sl Gauge theory correlators from non-critical string theory},
Phys. Lett. B 428 (1998) 105.

\bibitem{Witten}
E. Witten,
{\sl Anti-de Sitter space and holography},
Adv. Theor. Math. Phys. 2 (1998) 253.

\bibitem{Witten2}
E. Witten,
{\sl Anti-de Sitter space, thermal phase transition, and confinement in
gauge theories},
Adv. Theor. Math. Phys. 2 (1998) 505.

\bibitem{Duffads}
M. J. Duff,
{\sl Anti-de Sitter space, branes, singletons,
 superconformal field theories and all that},
{\tt hep-th/9808100}.

\bibitem{deWit1}
B.~de~Wit and H.~Nicolai,
\newblock {\sl $N=8$ supergravity with local $SO(8) \times SU(8)$ invariance},
\newblock Phys. Lett. 108B (1982) 285.

\bibitem{deWit2}
B.~de~Wit and H.~Nicolai,
\newblock {\sl $N=8$ supergravity},
\newblock Nucl. Phys. B 208 (1982) 323.

\bibitem{Hawkingturok}
N. Turok and S. W. Hawking,
{\sl Open inflation, the four-form and the cosmological constant},
Phys. Lett. B 432 (1998) 271.

\bibitem{Bousso}
R. Bousso and A. Chamblin,
{\sl Open inflation from nonsingular instantons: wrapping the
universe with a membrane},
{\tt hep-th/9805167}.

\bibitem{Bremer}
M. S. Bremer, M. J. Duff, H. Lu, C. N. Pope and K. S.
Stelle,
{\sl Instanton cosmology and domain walls from M-theory and string theory},
{\tt hep-th/9807051}.

\bibitem{Hawking}
S. W. Hawking, H. S. Reall,
{\sl Inflation, singular instantons and eleven-dimensional cosmology},
Phys. Rev. D 59 (1999) 023502.

\bibitem{Duffnilssonpope}
M. J. Duff, B. E. W. Nilsson and C. N. Pope,
{\sl Kaluza-Klein Supergravity},
Physics Reports 130 (1986) 1.

\bibitem{Klemm}
D. Klemm,
{\sl BPS black holes in gauged $N=4$ $D=4$ supergravity},
{\tt hep-th/9810090}.

\bibitem{Behrndt1}
K.~Behrndt, A.~H. Chamseddine and W.~A. Sabra,
\newblock {\sl BPS black holes in $N=2$ five-dimensional AdS supergravity},
Phys. Lett. B 442 (1998) 97.

\bibitem{Behrndt2}
K.~Behrndt, M. Cvetic and W.~A. Sabra,
\newblock {\sl Non-extreme black holes five-dimensional $N=2$ AdS
supergravity},
\newblock {\tt hep-th/9810227}.

\bibitem{Rahmfeld1}
M.~J. Duff and J.~Rahmfeld,
\newblock {\sl Massive string states as extreme black holes},
\newblock Phys. Lett. B 345 (1995) 441.

\bibitem{Kaluza}
M.~J. Duff,
\newblock {\sl Kaluza-{K}lein theory in perspective},
\newblock in {\em Proceedings of the Nobel Symposium {\it
Oskar Klein Centenary}, Stockholm, September 1994} (World Scientific, 1995),
E.~Lindstrom, editor,
\newblock {\tt hep-th/9410046}.

\bibitem{Khuri}
M. J. Duff, R.R. Khuri and J.X. Lu,
{\sl String solitons},
Phys. Rep. 259 (1995) 213.

\bibitem{Duffliurahmfeld1}
M.~J. Duff, J.~T. Liu and J.~Rahmfeld,
\newblock {\sl Four-dimensional string-string-string triality},
\newblock Nucl. Phys. B 459 (1996) 125.

\bibitem{Rahmfeld2}
J. Rahmfeld,
\newblock {\sl Extremal black holes as bound states},
\newblock  Phys. Lett. B 372 (1996) 198.

\bibitem{Rahmfeld3}
M.~J. Duff and J.~Rahmfeld,
\newblock {\sl Bound states of black holes and other $p$-branes},
\newblock  Nucl. Phys. B 481 (1996) 332.

\bibitem{Duffliurahmfeld2}
M.~J. Duff, J.~T. Liu and J.~Rahmfeld,
\newblock {\sl Dipole moments of black holes and string states},
\newblock Nucl. Phys. B 494 (1997) 161.

\bibitem{Biran}
B. Biran, A. Casher, F. Englert, M. Rooman, and P. Spindel,
{\sl The fluctuating seven-sphere in eleven-dimensional supergravity},
Phys. Lett. B 134 (1984) 179.

\bibitem{Sezgin}
E. Sezgin,
{\sl The spectrum of the eleven dimensional supergravity compactified
on the round seven sphere}, Trieste preprint, 1983, in Supergravity in
Diverse Dimensions, vol. 2, 1367, (eds A. Salam and E. Sezgin World
Scientific, 1989);
Fortschr. Phys. 34 (1986) 217.

\bibitem{Cremmer1}
E.~Cremmer and B.~Julia,
\newblock {\sl The $N=8$ supergravity theory. 1. The Lagrangian},
\newblock Phys. Lett. 80B (1978) 48.

\bibitem{Cremmer2}
E.~Cremmer and B.~Julia,
\newblock {\sl The $SO(8)$ supergravity},
\newblock Nucl. Phys. B 159 (1979) 141.

\bibitem{Romans}
L.~J. Romans,
\newblock {\sl Supersymmetric, cold and lukewarm black holes in cosmological
Einstein-Maxwell theory},
\newblock Nucl. Phys. B 383 (1992) 395.

\bibitem{Gibbons}
G.~W. Gibbons,
\newblock in  {\sl
Supersymmetry, Supergravity and Related Topics} (World Scientific, 1985),
F.~del Aguila, J.~A. Azcarraga and L.~E. Ibanez, editors.

\bibitem{Aichelburg}
P.~C. Aichelburg and F.~Embacher,
\newblock {\sl Exact superpartners of $N=2$ supergravity solitons},
\newblock Phys. Rev. D 34 (1986) 3006.

\bibitem{Lupope}
H. Lu and C. N. Pope,
{\sl P-brane solitons in maximal supergravities},
Nucl. Phys. B 465 (1996) 127.

\bibitem{Khuriortin}
R.~R. Khuri and T. Ortin,
\newblock {\sl Supersymmetric black holes in $N=8$ supergravity},
\newblock Nucl. Phys. B 467 (1996) 355.

\bibitem{Cvetictseytlin}
M. Cvetic and A. Tseytlin,
\newblock {\sl Solitonic strings and BPS saturated dyonic black holes},
\newblock  Phys. Rev. D 53 (1996) 5619.

\bibitem{Kalloshlinde}
R. Kallosh and A. Linde,
\newblock {\sl Supersymmetric balance of forces and condensation of
BPS states},
Phys. Rev. D 53 (1996) 5734.

\bibitem{Cvetic}
M.~{Cveti\v c} and D.~Youm,
\newblock {\sl Dyonic {BPS} saturated black holes of heterotic string on a
  six-torus},
\newblock Phys. Rev. D 53 (1996) 584.

\bibitem{dufflupope}
M. J. Duff, H. L\"u and C. N. Pope,
{\sl The black branes of M-theory}, Phys. Lett. B 382 (1996) 73.

\bibitem{Duffpope}
M. J. Duff and C. N. Pope,
{\sl Consistent truncations in Kaluza-Klein theories},
Nucl. Phys. B 255 (1985) 355.

\bibitem{Popetrunc}
C. N. Pope, {\sl Consistency of truncations in Kaluza-Klein},
in {\em Santa Fe 1984, Proceedings, The Santa Fe Meeting} (1985),
NSF-ITP-84-174.

\bibitem{Caldarelli}
M. M. Caldarelli and D. Klemm,
{\sl Supersymmetry of anti-de Sitter black holes},
{\tt hep-th/9808097}.

\bibitem{Freedmannicolai}
D. Z. Freedman and H. Nicolai,
{\sl Multiplet shortening in $OSp(N,4)$},
Nucl. Phys. B 237 (1984) 342.

\bibitem{Goddardnuytsolive}
P. Goddard, J. Nuyts and D. Olive,
{\sl Gauge theories and magnetic charge},
Nucl. Phys. B 125 (1977) 1.

\bibitem{GT}
G. W. Gibbons and P. K. Townsend,
{\sl Vacuum interpolation in supergravity via super $p$-branes},
Phys. Rev. Lett. 71 (1993) 3754.

\bibitem{DGT}
M. J. Duff, G. W. Gibbons and P. K. Townsend,
{\sl Macroscopic superstrings as interpolating solitons},
Phys. Lett. B 332 (1994) 321.

\bibitem{GHT}
G. W. Gibbons, G. T. Horowitz and P. K. Townsend,
{\sl Higher-dimensional resolution of dilatonic black hole
singularities},
Class. Quant. Grav. 12 (1995) 297.

\end{thebibliography}


\end{document}